\documentclass[printer]{aa}

\usepackage{graphicx}
\usepackage{txfonts}
\usepackage{epsf}
\usepackage{epsfig}

\def \etal{et~al.\/}
\def\lesssim{\mathrel{\hbox{\rlap{\hbox{\lower4pt\hbox{$\sim$}}}\hbox{$<$}}}}
\def\gtrsim{\mathrel{\hbox{\rlap{\hbox{\lower4pt\hbox{$\sim$}}}\hbox{$>$}}}}

\begin{document}

\title{Time-Dependent Behavior of Linear Polarization  in Unresolved Photospheres, With Applications for The Hanle Effect}

\author{R.~Ignace\inst{1}, K.~T.~Hole\inst{1}, J.~P.~Cassinelli\inst{2},
G.~D.~Henson\inst{1}
}

\institute{
    Department of Physics \& Astronomy, East Tennessee
    State University, Johnson City, TN, 37614 USA
    \and
    Department of Astronomy, 2535 Sterling Hall, University of Wisconsin,
        Madison, WI, 53706-1582 USA}

\offprints{ignace@etsu.edu}

\date{Received <date>; Accepted <date>}

\authorrunning{Ignace, Hole, Cassinelli, Henson}

\titlerunning{Time-Dependent Polarization in Unresolved Photospheres}

\abstract
{}
{
This paper extends previous studies in modeling time varying 
linear polarization due to axisymmetric magnetic fields in rotating stars. We use  the Hanle effect to predict variations in net line polarization, and use geometric arguments to generalize these results to linear polarization due to other mechanisms.
} 
{
Building on the work of Lopez Ariste \etal, we use simple
analytic models of rotating stars that are symmetric except for an axisymmetric magnetic field to predict the polarization lightcurve due to the Hanle effect.
 We highlight the effects for the
variable line polarization as a
function of viewing inclination and field axis obliquity. Finally, we use geometric arguments to generalize our results to linear polarization from the weak
transverse Zeeman effect.
  }
{We derive analytic expressions to demonstrate that the variable
polarization lightcurve for an oblique magnetic rotator is symmetric.
This holds for any axisymmetric field distribution and arbitrary
viewing inclination to the rotation axis.}
{For the situation under consideration, the amplitude of the
polarization variation is set by the Hanle effect, but the shape
of the variation in polarization with phase depends largely on
geometrical projection effects.  Our work generalizes the applicability of
results described in Lopez Ariste \etal, inasmuch as the assumptions
of a spherical star and an axisymmetric field are true, and provides
a strategy for separating the effects of perspective from the Hanle
effect itself for interpreting polarimetric lightcurves.  
}


\maketitle

\section{Introduction}

Polarization is a powerful tool in understanding diverse astrophysical
phenomena.  Net polarization in the continuum and spectral lines
can measure spatial, kinematic and compositional structure that
could not otherwise be detected in unresolved sources. It can also
be used as a diagnostic of magnetic fields in both resolved and
unresolved sources.  Continuum polarization can be produced by
Thomson scattering; in lines it can be created via the Hanle and
tranverse- and longitudinal-Zeeman effects. In this paper we focus
on the Hanle effect, but our results can also apply to other
magnetically-induced linear polarization mechanisms.

The Hanle effect is a weak field case of the Zeeman effect with
consequence for resonance line scattering polarization in the
presence of a magnetic field (e.g., Moruzzi \& Strumia 1991; Stenflo
1994; Landi Degl' Innocenti \& Landolfi 2004).  The Hanle effect describes the
influence of the field for the {\em linear} polarization of line scattering;
unlike the Zeeman effect, the Hanle effect does not generate circularly
polarized emissions.
%
%

The Hanle effect has proven to have
diagnostic value in a number of applications to solar physics.  Examples
from the recent literature include coronal
magnetic field (e.g., Derouich \etal\ 2010), turbulent magnetic fields
(e.g., Frisch \etal\ 2009; Rachkovskii 2009; Kleint \etal\ 2010),
chromospheric magnetic fields (e.g., Faurobert \etal\ 2009), and prominences
(e.g., Merenda \etal\ 2006), to name only a few.  There have also been
efforts to develop diagnostics based on the Hanle effect for molecular
lines (e.g., Berdyugina \& Fluri 2004; Shapiro \etal\ 2007).  Many
of these diagnostics have been developed to interpret the so-called
``Second Solar Spectrum'' (Stenflo \& Keller 1997).  There has even
been a consideration of the Hanle effect for the magnetic field
of Jupiter (Ben-Jaffel
\etal\ 2005).

For stars other than the Sun, considerations of the Hanle effect
have so far been restricted to theoretical calculations, such as
simplified considerations in stellar wind lines (e.g., Ignace,
Nordsieck, \& Cassinelli 2004), circumstellar disks (Yan \& Lazarian
2008; Ignace 2010), and maser sources (Asenio Ramos, Landi
Degl'Innocenti, \& Trujilo Bueno 2005).  So far, no definitive
detections of the Hanle effect in the stellar context have been
reported.  

A new consideration of the Hanle effect in photospheric lines of
unresolved stars has been proposed by Lopez Ariste, Asensio Ramos, \&
Gonzalez Fernandez (2011).  Our paper makes a contribution to these new
and interesting results by extending them to a more general case through
a consideration of perspective effects for oblique magnetic rotators.
We demonstrate that for a given field topology and field strength, aspects
of the variable polarization due to geometry can be disentangled from
the mechanism of the Hanle effect and generalized to apply to any net
polarization due to a bipolar magnetic field.  Section~\ref{sec:hanle}
presents the background for our models: section~\ref{sub:pol} describes
polarization conventions and gives an overview of the nature of
the Hanle effect.  Adopted geometry and assumptions are defined in
section~\ref{sub:conventions}, and solutions for perspective effects
for polarimetric lightcurves are given in section~\ref{sub:solns}.
There are three parts to the discussion:  (a) a consideration of an
edge-on (or equator-on) rotating star as discussed by Lopez Ariste \etal\
(2011); (b) a generalization to arbitrary viewing inclination of the
stellar rotation axis; and (c) a review of a special case in which the
dependence of amplitude on the Hanle effect is a known function of the
inclination between the field and observer axes.  Concluding remarks
are given in section~\ref{sec:concs}. 
An extra figure, exclusive to the {\em astro-ph} version of this paper, is included in the Appendix.

\section{The Hanle Effect in Unresolved Photospheres}	\label{sec:hanle}

\subsection{Polarization and the Hanle Effect}	
\label{sub:pol}

Polarization is measured in terms of the four Stokes vector components,
$I$, $Q$, $U$ and $V$.  Here $I$ is total intensity; $Q$ and $U$
measure the intensity of linearly polarized light relative to two
axes offset by a rotation of $45^\circ$; and $V$ is a measure of
the circularly polarized light -- in our case zero. For convenience we 
introduce normalized parameters $q=Q/I$ and $u=U/I$.

Measured Stokes parameters are defined with respect to one orientation
in the sky -- by convention, $q$ is measured along the North-South
axis.  Of course, some arbitrary source in the sky will not normally
have a favorable orientation with respect to the observer convention.
In addition to measuring polarized flux in the Stokes 4 parameter
system, it is useful to have a relative measure of the polarization
that is independent of the observer system.

With a focus on the linear polarization, the observer-dependent $q$
and $u$ values can be recast in terms of a polarization magnitude
and orientation on the sky. These are $p$, the fraction (or percentage)
of polarized light, and $\psi$, the position angle. They are defined
as

\begin{equation} 
p =\sqrt{q^2+u^2},
\end{equation}

\noindent and 

\begin{equation}
\tan 2\psi = \frac{U}{Q}= \frac{u}{q}.
\end{equation}

\noindent These definitions are crucial to arguments that will
appear in the following sections.  Measures of polarization using
$(q,u)$ and $(p,\psi)$ are operationally analogous to Cartesian and
polar coordinates.  In this respect $p$ acts like a radius.  A
rotation of the coordinate frame to a new system affects the $(q,u)$
measures, but {\em not} $p$.

The Hanle effect can produce a net polarization in the line
in the following way.
First consider a spherically symmetric star that has no magnetic
field.  The polarization profile is centro-symmetric, so that as
an unresolved source, the star has net zero polarization.  The
radial profile of the polarization is zero at the projected center
of the star, and has a maximum value denoted by $p_0$ at the stellar
limb with an orientation tangent to the limb.  The value of $p_0$
can vary from one line to the next.  Indeed, even the shape of the
run of polarization from center to limb can differ between lines,
but this will not affect our conclusions.

The presence of a magnetic field breaks the symmetry across the
projected stellar disk leading to a net polarization in the line.
A semi-classical description explains the Hanle effect in terms
of a damped harmonic oscillator.  The magnetic field serves to
precess the orientation of the oscillator, leading to a change in
polarization position angle and amplitude relative to the zero-field
case (Hanle 1924).  As a result, a net polarization in an unresolved
line from an unresolved photosphere can be produced that is a
function of the field geometry, field strength, and the line
transition under consideration.  The analysis that follows emphasizes
geometrical considerations and symmetries for the case of polarized
lines from oblique magnetic rotators.

\subsection{Conventions and Assumptions}	
\label{sub:conventions}

To investigate the observational implications of magnetic-field-induced
polarization in rotating stars, we make a number of reasonable
simplifying assumptions. (See Figure~\ref{fig1} for the definition
of our geometry.)

\begin{itemize}

\item Except for the field distribution, the star is otherwise
spherically symmetric.

\begin{figure}[t]
\resizebox{\hsize}{!}{\includegraphics{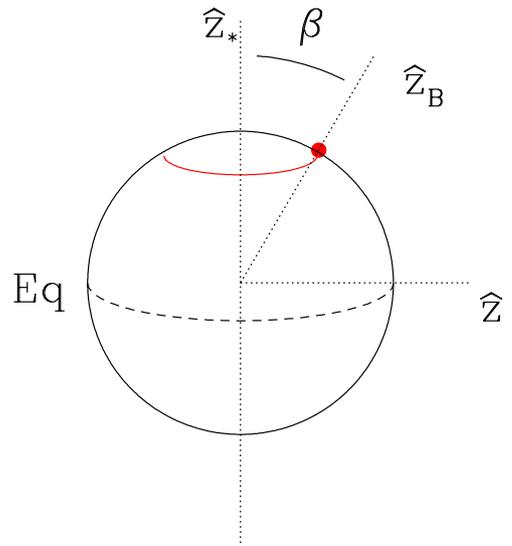}}
\caption{Illustration of the relation between the rotation, magnetic,
and observer axes ($z_\ast$, $z_B$, and $z$ respectively) as
seen equator-on.  The
dashed line is the equator and the red arc is the path followed
by the magnetic pole as the star rotates.  The red dot is
the intersection of the magnetic axis with the stellar surface
(compare with Fig.~\ref{fig2}).  In this figure the magnetic
pole is currently at longitude of zero (i.e., $\phi=0^\circ$).
\label{fig1}}
\end{figure}

\begin{figure}[t]
\resizebox{\hsize}{!}{\includegraphics{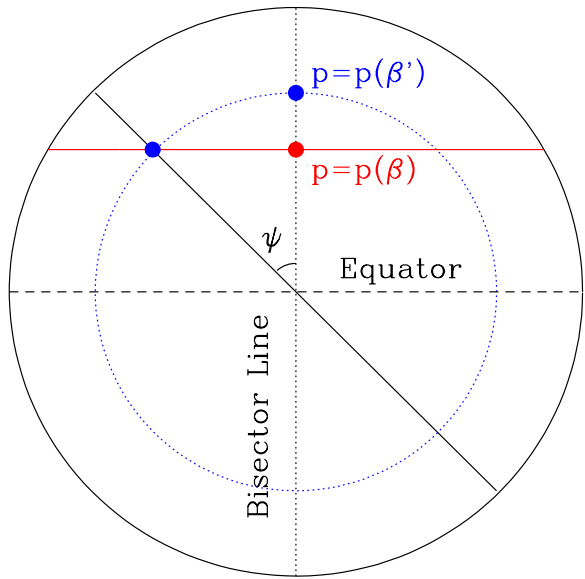}}
\caption{Illustration of the projected stellar disk for a star
seen equator-on to its rotation.  The equator is indicated.  The
``bisector line'' defines zero longitude ($\phi=0^\circ$) on the star.  The red
line represents the constant co-latitude of rotation for the
magnetic axis at the point of interception with the stellar 
surface.  The blue dotted circle indicates the locus of points which
would have a constant polarization $p$ if the magnetic pole were 
located on that curve, with the two blue dots as examples.
\label{fig2}}
\end{figure}

\item The star is taken to rotate as a solid body.  The observer
is located in direction $\hat{z}$ from the center of the star.  The
rotation axis is in direction $\hat{z}_\ast$.  The viewing inclination
$i_0$ is defined by

\begin{equation}
\hat{z}\cdot\hat{z}_\ast = \cos i_0.
\end{equation}

\item The magnetic field is dipolar and axisymmetric.  Its symmetry axis is
in direction $\hat{z}_B$.  The obliquity angle $\beta$ between
the field axis and the rotation axis is defined by 

\begin{equation}
\hat{z}_\ast \cdot \hat{z}_B = \cos \beta.
\end{equation}

\item 
We adopt the definition of stellar longitude introduced by Lopez Ariste \etal, using the angle $\phi$ to
represent the longitude of the field axis on the stellar surface.
(The location of this axis is of course simply a point.)  Zero longitude ($\phi=0^\circ$) means
that the magnetic north pole lies in front of the star along the
bisector (or ``Prime Meridian'') of the observer's projected stellar disk in the sky.

\end{itemize}

Figure~\ref{fig2} gives an illustration of the projected stellar
disk.  The dashed line is the equator; dotted is the
bisector line with $\phi=0^\circ$.  The red dot
represents the location of the intersection of the magnetic axis
(which passes through the star's center) with the stellar surface.
This axis is inclined by the angle $\beta$ from the north pole (top
of the circle).  

We use this bisector line to define observer axes for measured
Stokes parameters $I, Q, U,$ and $V$.  Circular polarization is not
relevant to the discussion, so $V=0$.  The angle $\psi$ is the
reference orientation angle for the field axis on the stellar surface
as measured counterclockwise around the observer axis from the
bisector line.  Stokes $Q$ and $U$ are defined in terms of the net
surface integrated polarized flux arising from the Hanle effect and
in reference to this orientation angle.  Tranformation of our $Q$
and $U$ values to some other observer system $Q'$ and $U'$ is easily
accomplished by means of a rotation through an angle $2\psi_0$,
viz.

\begin{equation}
\left(\begin{array}{c}
Q' \\ U'
\end{array}\right) = \left( \begin{array}{cc}
\phantom{-}\cos 2\psi_0 & \sin 2\psi_0 \\
-\sin 2\psi_0 & \cos 2\psi_0 \end{array} \right)
\,\left(\begin{array}{c}
Q \\ U
\end{array}\right),
\end{equation}

\noindent The point of going through the exercise of the preceding
expression is that net values of $Q$ and $U$ in the line will most
easily be evaluated in a reference system defined by the field axis
because of the underlying physics, but polarized fluxes will be
measured in an observer system (the primed system).  The quantity
that is preserved in this transformation is the polarization
$p=\sqrt{q^2+u^2}=\sqrt{q'^2+u'^2}$.

We stress that, as in Lopez Ariste \etal\ (2011), the polarization
under consideration is the {\em total} line polarization.  We are not
discussing temporal behavior within resolved lines.

\subsection{Analytic Solutions}	\label{sub:solns}

We take as given that solutions for $p$ as a function of field
strength $B$, obliquity $\beta$, and viewing inclination of the
rotation axis $i_0$ have already been found for the line transition
under consideration, for example from calculations like those given
in Lopez Ariste \etal\ (2011; see their Figs.~1 and~2).  That paper
presents a lightcurve in their Figure~3 that shows an asymmetric
variable polarization with respect to zero longitude, corresponding
to our bisector line with $\phi=0^\circ$.  We believe,
however, that the lightcurve should be mirror symmetric about this
reference line, for reasons presented in the next section.  Finally,
we develop an extension of that conclusion in relation to general
axisymmetric field distributions for rotating stars viewed at
arbitrary inclinations.  We then discuss a particular application
that has been presented in Ignace (2001).

\subsubsection{Symmetry of the Polarization Lightcurve}	\label{subsub:sym}

Let the rotating star be viewed equator-on, with $i_0 = 90^\circ$.
Suppose that an ensemble of calculations provide the line polarizations
for this case.  As the star in Figure~\ref{fig2} rotates, the red
dot would track along the red line.  When it passes behind the star,
the opposite magnetic pole would appear in the southern hemisphere
in a mirror-symmetric way.

The blue dot that lies on the red line is the projected location
of the magnetic axis at some time during the rotation.  For the
observer that dot (the magnetic pole)
is at an orientation of $\psi$; in the star system,
the dot is located at longitude $\phi$.  Recall that the star is
spherically symmetric, the field is axisymmetric, and the photodisk
is unresolved.  This means the total observed polarization from the
star must correlate with the location of the magnetic pole and be
a periodic function of rotational phase.  We refer to this polarization
as $p(90^\circ,\beta,B,\phi)$.

An efficient way to think of the polarized lightcurves is to employ
the fact that polarization $p$ is invariant with rotation of the
observer reference frame.  If one were to take the blue dot on the
red line and simply rotate it about the observer axis, it would
trace out the dotted blue circle shown in Figure~\ref{fig2}.  Remember,
the bisector line in Figure~\ref{fig2} is defined in relation to the
star's rotation axis and cannot generally be expected to line up
with the North-South direction at Earth!  Will the failure of these
two to line up change the observed polarization?  Of course not, because
$p$ is invariant under such a rotation.

Therefore, rotating a field of the same topology, strength, and
projection as signified by the blue dot on the red line to any other
position on the dotted blue circle will not change the value of
$p$.  The blue circle intersects the bisector line.  We see then
that the polarization at any rotational longitude (or rotational
phase) is equivalent to the polarization value on the bisector line
for a different field obliquity $\beta'$.

This graphical construction leads to the following relation:

\begin{equation}
p(90^\circ,\beta,B,\phi) = p(90^\circ, {\beta}',B,0).
\label{eq:edgeon}
\end{equation}

\noindent Simple spherical trigonometry gives

\begin{eqnarray}
\sin \beta' & = & \sin \beta \, \cos \phi,\;{\rm and} \\
\tan \psi & = & - \frac{\tan \phi}{\cos\beta} \, .
\label{eq:edgetrig}
\end{eqnarray}

\noindent As expected, $\beta'=\beta$ and $\psi=0^\circ$ when the
field axis is oriented along zero longitude of $\phi=0^\circ$, the
bisector line.

In this edge-on scenario, knowing the polarization from models as
a function of line transition, field strength, and (axisymmetric)
field topology along the bisector line alone allows one to construct
polarized lightcurves for oblique magnetic rotators.  Variations
in $q$ and $u$ will be given by:

\begin{eqnarray}
q & = & p(90^\circ,\beta',B,0^\circ)\,\cos(2\psi),\\
u & = & p(90^\circ,\beta',B,0^\circ)\,\sin(2\psi).
\end{eqnarray}

\noindent The relations of equation~(\ref{eq:edgeon})--(\ref{eq:edgetrig})
demonstrate that $\beta'$ takes on the same value at $\phi$ as for
$-\phi$.  Therefore the polarimetric lightcurves must be symmetric
about zero longitude.

\subsubsection{Generalization to Arbitrary Inclinations}  \label{subsub:arb}

\begin{figure}[t]
\resizebox{\hsize}{!}{\includegraphics{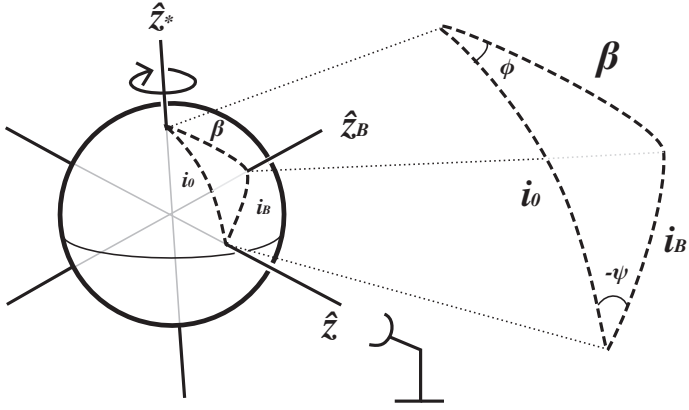}}
\caption{The three dimensional geometry for the general case of
arbitrary inclination of the rotation axis. The close-up shows
details of the spherical triangle defined by the three axes in the
problem, those  of the observer, the star's rotation, and the
magnetic field.  Symbols are defined in the text.  Axes
meet at the center of the star, but are not necessarily
perpendicular to each other. \label{fig3}}
\end{figure}

Figure~\ref{fig3} illustrates the spherical geometry in the general
case of an arbitrary viewing inclination.  The direction of the
rotation axis is $\hat{z}_\ast$.  The observer is located in the
direction of $\hat{z}$ in this figure, and the magnetic axis in the
direction of $\hat{z}_B$.  The angle $i_B$ represents the effective
viewing inclination of the field axis at any moment.  Note that
$\phi$ is measured from the plane defined by the observer and
rotation axes as measured counterclockwise around $z_\ast$.  The
observer orientation $\psi$ is similarly defined in relation to the
$z$-axis.  In the spherical triangle of Figure~\ref{fig3}, $-\psi$
appears because the magnetic pole lies clockwise of the reference
plane.

For purposes of generalization, the argument of the previous section
requires some new considerations.  If the rotation axis is not seen
equator-on, it might not be the case that the polarization is
invariant for the field axis located anywhere around the blue curve.
(Note that the pole-on case is trivial, with
no variation in polarization.)  But consider the following.

\begin{itemize}

\item If the field is axisymmetric and top-down symmetric,
then $p$ is invariant around the blue circle.  

\item If the field is axisymmetric, but not top-down symmetric,
meaning it is not mirror symmetric about the magnetic equator, then
$p$ is not invariant around the blue circle.  When the field axis
(i.e., the red dot in Fig.~\ref{fig2}) is at $\phi< -90^\circ$ or
$\phi>+90^\circ$, the observer sees more of the southern magnetic
hemisphere than the northern, which is the opposite of the case
when $\phi>-90^\circ$ and $\phi<90^\circ$.  However, it remains the
case that the lightcurve is symmetric about zero longitude.

\end{itemize}

Here a comment regarding the meaning of ``top-down symmetric'' is
needed.  The divergence of a magnetic field is zero.  The implication
for top-down symmetry is that the field topology above and below
the magnetic equator is identical except for a vector reversal of
the field, meaning $\vec{B}_{\rm North} = -\vec{B}_{\rm South}$ for
the poloidal component of the field.

Technically, this kind of symmetry
would matter for the Hanle effect.  A sign flip
of the vector field does change $Q$ and $U$ qualitatively: the
precession of the oscillating charge in the semi-classical description
of the Hanle effect now proceeds with opposite handedness.  But at
the same field strength, the \emph{total} polarization is not
changed.  In our ``top-down'' symmetry, a change in the sign of the
vector field does not affect the amplitude of the polarization.

To summarize, the Hanle effect modifies the polarized surface map
across a star.  That map depends on three things: (a) the field
topology, (b) the field strength, and (c) the orientation of the
field with respect to the observer.  However, the star is not
resolved, so a given polarimetric map from a ``snapshot'' of the
star reduces to a single value of $q$ and $u$, or equivalently $p$
and $\psi$.  If the field is axisymmetric and top-down symmetric,
then for a given field topology and field strength, the value of
the net polarization $p$ tracks with the projected location of the
field symmetry axis on the star only.  This means $p$ is a constant
for an axis location on circles of constant projected-disk radius
centered on the observer's line-of-sight to the star center.

As a result of these arguments, $p=p(\{\hat{B}\}, B, i_B)$, where
$\{\hat{B}\}$ represents the field topology at the location of line formation,
where $B$ again is the field strength, and now $i_B$ is the
instantaneous angle between the observer and the field axis, given
by $\cos i_B = \hat{z}_B\cdot \hat{z}$.  Knowing the polarization
for all $i_B$ amounts to knowing the polarization for a magnetic
axis at any location on the projected disk.  Schematically,
polarimetric lightcurves can be constructed by simply considering
how the point of the magnetic pole on the star moves through this
grid of known solutions for $p$ as a function of longitude $\phi$.

The time behavior of the polarization is cyclic, symmetric, and
governed by how the magnetic pole tracks across the face of the
projected stellar disk which is set by the angles $i_B$ and $\psi$.
Working out the spherical trigonometry based on Figure~\ref{fig3},
solutions for $i_B$ and $\psi$ in terms of field axis obliquity,
rotation axis inclination, and longitude of the magnetic pole are
given by

\begin{eqnarray}
\cos i_B & = & \cos\beta\,\cos i_0+\sin \beta\,\sin i_0\,\cos \phi, \\
\tan\psi & = & -\frac{\sin\beta\,\sin i_0\,\sin\phi}
	{\cos\beta\,\sin i_0 + \sin \beta\, \cos i_0\,\cos\phi}.
\end{eqnarray}

\noindent The $q$ and $u$ Stokes parameters at any time
(or phase) in the lightcurve will now be given by

\begin{eqnarray}
q & = & p(\{\hat{B}\},B,i_B)\,\cos(2\psi), \label{eq:q}\\
u & = & p(\{\hat{B}\},B,i_B)\,\sin(2\psi). \label{eq:u}
\end{eqnarray}

\noindent where $i_B$ and $\psi$ are implicit functions of
the longitude $\phi$.  

The top-down symmetry as we have described ensures that $p$ is a
constant regardless of which end of the magnetic field axis appears
on the projected stellar disk.  For certain values of $i_0$ and
$\beta$, the north magnetic pole can rotate behind the star and be
eclipsed.  At those moments of ingress and egress, the south magnetic
pole will be at egress or ingress, respectively.  Although $p$ does not
change, $\psi$ tracks with the location of the field axis and thus
suffers a $180^\circ$ rotation.  However, this will produce no
observable effect since $q$ and $u$ depend on $2\psi$, and so
are periodic with $180^\circ$.

One interesting note is that the ratio of $u/q$ (and thus $\psi$)
is completely independent of the Hanle effect and
derives purely from geometrical considerations of the problem.  The
ratio is

\begin{equation}
\frac{u}{q} = \tan(2\psi) = \frac{2\tan \psi}{1-\tan^2\psi},
\end{equation}

\noindent and contains information about $i_0$ and $\beta$
only.  This result thus provides an interesting control in the modeling
effort of observations by relegating the influence of the Hanle
effect to the shape of the total polarization lightcurve.

\subsubsection{Special Case of Separable Functionality}  \label{subsub:special}

Without knowing the exact dependence of the polarization with angle
$i_B$, it is not possible to evaluate $q$ and $u$ any further
than expressed in equation~(\ref{eq:q})--(\ref{eq:u}).  However, there
is at least one case where that dependence is known.  

Ignace, Nordsieck, \& Cassinelli (1997) considered the Hanle effect
for optically thin resonance line scattering in circumstellar media,
such as winds and disks.  For a spherically symmetric wind and an
axisymmetric field topology (analogous to the photospheric case
considered by Lopez Ariste \etal\ 2011), Ignace \etal\ (1997) showed
that the net line polarization scales as $p \propto f(\vec{B},n)\,\sin^2
i_B$.  Here $n$ relates to the spherical number density distribution.
The function $f$ depends on the density (in effect, the line optical
depth), the field strength and topology, and the particular line
transition.  Ignace (2001) then considered the Hanle effect from
oblique magnetic rotators for emission lines that form in a spherically
symmetric circumstellar environment.

Although the adoption of spherical symmetry in the circumstellar
environment is not a physically realistic scenario, it does serve
as a concrete example in which the conclusions of the preceding
sections can be illustrated.  The factor $\sin^2 i_B$ is a particular
solution for thin lines when treating the star as a point source
in terms of illumination and occultation.  Under these circumstances,
the Hanle effect appears only as a scale parameter that sets the
amplitude of the polarization completely isolated from the $i_B$-dependence.
Thus, polarimetric variations arise strictly from geometrical
considerations in $q/f$ and $u/f$, as displayed in Figure~2 of 
Ignace (2001).

\section{Conclusions}	\label{sec:concs}

We have shown that the polarimetric lightcurves arising from the
Hanle effect in photospheric lines of oblique magnetic rotators are
mirror symmetric in relation to a stellar-disk ``bisector'' line.
We have additionally described a methodology extending the interesting
results presented in Lopez Ariste \etal\ (2011) for the equator-on
case to more general viewing inclinations.  We have further shown
explicitly that the ratio $u/q$ is independent of the Hanle effect
and varies with time purely as a function of perspective variables
for the magnetic axis.

In fact the geometric properties that we have employed in relation
the lightcurves for Hanle polarization have more general applicability.
Fox (1992) showed applicability of this approach to polarizations
arising from electron scattering in stellar envelopes.  
This has been used
in simulating variable polarizations from clumped wind
flows.  Davies, Vink \& Oudmaijer (2007) and Li \etal\ (2009) derive
the polarization from an individual electron scattering wind clump
and then use the geometric properties emphasized in this paper to
``populate'' a wind with many clumps to explain observed variable
polarizations in Luminous Blue Variable stars and Wolf-Rayet stars.

Moreover, the results should hold for the
total linear polarization of lines in which the transverse Zeeman
effect is operating.  Indeed, Landolfi \etal\ (1993) demonstrate
this explicitly in the ``weak'' limit of the transverse Zeeman
effect.  For an oblique magnetic rotator with a dipole field,
equation~14 and Figure~2 of Landolfi \etal\ (1993) are to be compared
with (as it happens) Figure~2 for the special case noted above due to
the Hanle effect in Ignace \etal\ (2001).  The two are in fact the
same in terms of geometrical and time variable dependence.

The end result is that we find aspects of geometry that decouple
from the physical mechanisms that give rise to observed variable
polarimetric signals.  This is important for distinguishing geometrical
effects from source properties.  Furthermore in the case of the
Hanle effect, a diagnostic approach that evaluates polarizations
in {\em different} lines will prove important.  This is because for
a given field strength, different lines have different levels of
response to the Hanle effect.  Ignace \etal\ (1997) discussed the
power of a multi-line approach for discerning circumstellar magnetic
fields from polarized line data.  Although they dealt with a
steady-state scenario, the advantages that come from a multiline
analysis will be useful in deriving Hanle effect solutions for field
geometries and strengths in oblique magnetic rotators.

\begin{acknowledgements}

The authors are indebted to the referee V.~Bommier for a thorough review
of this manuscript.  We gratefully acknowledge funding for this work
provided by the National Science Foundation, grant AST-0807664.

\end{acknowledgements}

\onecolumn

\begin{appendix}

\section{Bonus Material}    \label{sec:extra}

\begin{figure*}[htbp]
\begin{center} 
\includegraphics[width=6.5in]{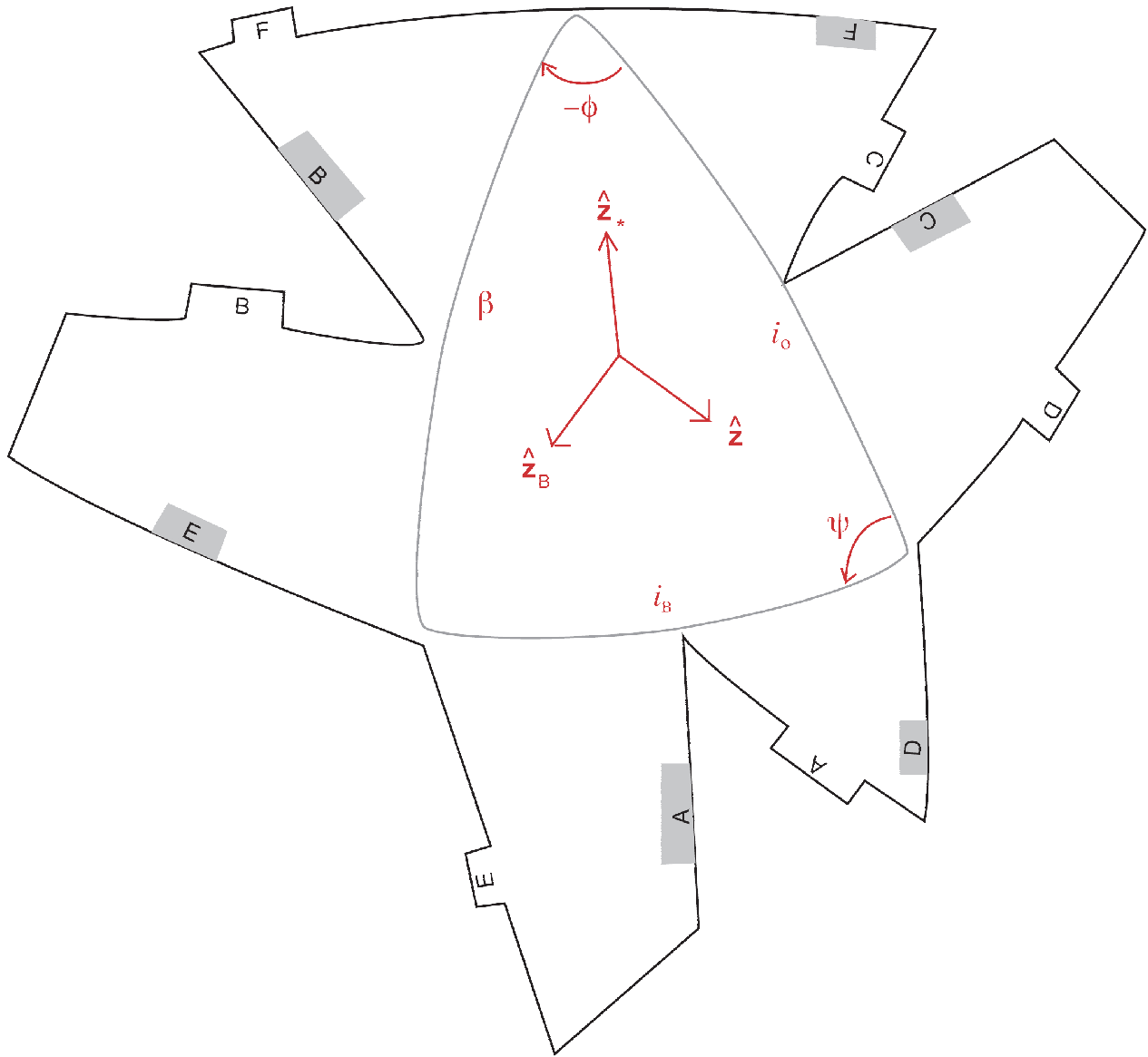}
\caption{{\bf Three-dimensional spherical trigonometry model} -- With this figure, you can create your own 3-D sector of a sphere, to help visualize the geometries involved in this paper. Just cut out the above figure along the black lines, then tape the sides together so that Tab A is over the gray box labeled A, Tab B over Box B, through Tab F and Box F. When it is assembled, the gray line will demarcate an area on the surface of the star. The angles, sides and axes used in the paper are labeled on this surface in red.  }
\label{f:bonus}
\end{center}
\end{figure*}

This appendix presents an extra figure exclusive to the astro-ph version of this paper. Figure \ref{f:bonus} can be cut out and assembled to create a model of a sector of the star. It may also be useful as an ornament for any festive occasion, or at least those involving astronomers.

\vspace{0.5in}

\end{appendix}


\begin{thebibliography}{}

\bibitem {} Asenio Ramos, A., Landi Degl'Innocenti, E., Trujillo
	Bueno, J., 2005, ApJ, 625, 985
\bibitem {} Ben-Jaffel, L, Harris, W., Bommier, V., Roesler, F., 
	Ballester, G.~E., Jossang, J., 2005, Icar, 178, 297
\bibitem {} Berdyugina, S.~V., Fluri, D.~M., 2004, A\&A, 417, 775
\bibitem {} Davies, B., Vink, J.~S., Oudmaijer, R.~D., 2007, A\&A,
	469, 1045
\bibitem {} Derouich, M., Auchere, F., Vial, J.~C., Zhang, M., 2010,
	A\&A, 511, 7
\bibitem {} Faurobert, M., Derouich, M., Bommier, V., Arnaud, J.,
	2009, A\&A, 493, 201
\bibitem {} Fox, G.~K., 1992, Ap\&SS, 187, 219
\bibitem {} Frisch, H., Anusha, L.~S., Sampoorna, M., Nagendra, K.~N.,
	2009, A\&A, 501, 335
\bibitem {} Hanle, W., 1924, Z.\ Phys., 30, 93
\bibitem {} Ignace, R., 2001, in Advanced Solar Polarimetry, M.~Sigwarth
	(ed.), ASP Conf.\ Ser.\ \#236, (ASP:  San Francisco), 227
\bibitem {} Ignace, R., 2010, ApJ, 725, 1040
\bibitem {} Ignace, R., Nordsieck, K.~N., Cassinelli, J.~P., 1997,
	ApJ, 486, 550
\bibitem {} Ignace, R., Nordsieck, K.~N., Cassinelli, J.~P., 2004,
	ApJ, 609, 1018
\bibitem {} Kleint, L., Berdyugina, S.~V., Shapiro, A.~I., Bianda,
	M., 2010, A\&A, 524, 37
\bibitem {} Landi Degl'Innocenti, E., Landolfi, M., 2004, Polarization in
	Spectral Lines, (Kluwer:  Dordrecht)
\bibitem {} Landolfi, M., Landi Degl'Innocenti, E., Landi Degl'Innocenti,
	M., Leroy, J.~L., 1993, A\&A, 272, 285
\bibitem {} Li, Q.-K., Cassinelli, J.~P., Brown, J.~C., Ignace, R.,
	2009, RAA, 9, 558
\bibitem {} Lopez Ariste, A., Asensio Ramos, A., Gonzalez Fernandez, C.,
	2011, A\&A, 527, A120
\bibitem {} Merenda, L., Trujillo Bueno, j., Landi Degl'Innocenti, E.,
	Collados, M., 2006, ApJ, 642, 554
\bibitem {} Moruzzi, G., Strumia, F., (eds.), 
	1991, The Hanle Effect and Level-Crossing Spectroscopy, (Plenum Press:
	New York)
\bibitem {} Rachkovskii, D.~N., 2009, Astronomy Reports, 53, 153
\bibitem {} Shapiro, A.~I., Berdyugina, S.~V., Fluri, D.~M., Stenflo,
	J.~O., 2007, A\&A, 475, 349
\bibitem {} Stenflo, J.~O., Solar Magnetic Fields, (Kluwer:  Dordrecht),
	1994
\bibitem {} Stenflo, J.~O., Keller, C.~U., 1997, A\&A, 321, 927
\bibitem {} Yan, H., Lazarian, A., 2008, ApJ, 677, 1401

\end{thebibliography}
\end{document}